\documentclass{article}
\usepackage{graphicx} 
\usepackage{authblk}
\usepackage[a4paper, total={6in, 9in}]{geometry}
\usepackage{hyperref}
\usepackage{amsmath}
\usepackage{amssymb}

\title{The serotonergic psychedelic N,N-dipropyltryptamine alters information-processing dynamics in cortical neural circuits.}

\author[1,2]{Thomas F. Varley}
\author[3]{Daniel Havert}
\author[3]{Leandro Fosque}
\author[2,4]{Abolfazl Alipour}
\author[3]{Naruepon Weerawongphrom}
\author[5]{Hiroki Naganobori}
\author[5]{Lily O'Shea}
\author[1,4]{Maria Pope}
\author[3,4]{John Beggs}

\affil[1]{School of Informatics, Computing, \& Engineering, Indiana University, Bloomington, IN, USA}
\affil[2]{Department of Psychological \& Brain Sciences, Indiana University, Bloomington, IN, USA}
\affil[3]{Department of Physics, Indiana University, Bloomington, IN, USA}
\affil[4]{Program in Neuroscience, Indiana University, Bloomington, IN, USA}
\affil[5]{Indiana University, Bloomington, IN, USA}

\date{July 2023}

\begin{document}

\maketitle

\begin{abstract}
    Most of the recent work in psychedelic neuroscience has been done using non-invasive neuroimaging, with data recorded from the brains of adult volunteers under the influence of a variety of drugs. While this data provides holistic insights into the effects of psychedelics on whole-brain dynamics, the effects of psychedelics on the meso-scale dynamics of cortical circuits remains much less explored. Here, we report the effects of the serotonergic psychedelic N,N-diproptyltryptamine (DPT) on information-processing dynamics in a sample of \textit{in vitro} organotypic cultures made from rat cortical tissue. Three hours of spontaneous activity were recorded: an hour of pre-drug control, and hour of exposure to 10$\mu$M DPT solution, and a final hour of washout, once again under control conditions. We found that DPT reversibly alters information dynamics in multiple ways: first, the DPT condition was associated with higher entropy of spontaneous firing activity and reduced the amount of time information was stored in individual neurons. Second, DPT also reduced the reversibility of neural activity, increasing the entropy produced and suggesting a drive away from equilibrium. Third, DPT altered the structure of neuronal circuits, decreasing the overall information flow coming into each neuron, but increasing the number of weak connections, creating a dynamic that combines elements of integration and disintegration. Finally, DPT decreased the higher-order statistical synergy present in sets of three neurons. Collectively, these results paint a complex picture of how psychedelics regulate information processing in meso-scale cortical tissue. Implications for existing hypotheses of psychedelic action, such as the Entropic Brain Hypothesis, are discussed. 
\end{abstract}

\section{Introduction}

Serotonergic psychedelics such as LSD, psilocybin, and mescaline, are known to induce intense, exotic states of consciousness that depart markedly from normal day-to-day patterns of cognition and perception \cite{nichols_psychedelics_2016}. Since the turn of the century, there has been a resurgence of interest in the scientific exploration of psychedelic states, with a particular focus on using whole-brain neuroimaging technologies to understand the neural correlates of the psychedelic experience. In typical recent studies, adult human volunteers are given a psychedelic, and then brain activity is recorded for analysis, which can then be compared to self-reported phenomonological experiences (such as the experience of ego dissolution \cite{lebedev_finding_2015}), or clinical presentations (such as depression \cite{carhart-harris_psilocybin_2017}). Human neuroimaging studies have been done using almost every available modality, including fMRI (for a review of existing fMRI dataset, see \cite{mcculloch_psychedelic_2022}), EEG (for a partial review of EEG studies, see \cite{tofoli_chapter_2016}) and MEG \cite{carhart-harris_neural_2016}. Collectively, these studies have painted a complex picture of the effects of different psychedelics on whole-brain, macro-scale activity, with one of the most-discussed effects being a general increase in the entropy (or ``complexity") of macro-scale brain activity (for review, see \cite{sarasso_consciousness_2021}, although for an recent study into which specific measures of entropy replicate, see \cite{mcculloch_navigating_2023}). This apparent link prompted Carhart-Harris and colleagues to propose the so-called ``entropic brain hypothesis" (EBH), which posits a link between the information density of spontaneous brain activity and the perceptual richness or lability of conscious experience \cite{carhart-harris_entropic_2014,carhart-harris_entropic_2018}. 

There have been far fewer attempts to understand the micro-scale, circuit-level effects of psychedelics. This creates something of a schism in the field of psychedelic science: at the level of individual neurons, ligands, and receptors, the pharmacological properties of psychedelics are well understood \cite{nichols_psychedelics_2016}, and at the level of the entire brain, the effects of psychedelics on brain dynamics are beginning to crystallize as well (increased complexity of spontaneous activity, etc \cite{carhart-harris_entropic_2018,sarasso_consciousness_2021}). However, the intermediary circuit-level dynamics induced by psychedelics at the ``meso-scale", which presumably form the causal substrate of the high-level dynamical changes, remains largely unexplored. 

The few studies that have been done in this space have largely focused on single measures, such as firing rate \cite{wood_disruption_2012,rajpal_quantifying_2023}, or coherence \cite{brys_classic_2022}. Our goal with this study was a more comprehensive analysis of how a serotonergic psychedelic alters the information-processing dynamics of neural circuits. Information dynamics \cite{lizier_local_2013} is a branch of information theory concerned with the understanding how distributed systems ``compute" their trajectories through configuration space over time.  Prior work has shown that the information dynamics framework applied to spiking neural activity is powerful enough to reveal meaningful differences in cognitive state and behaviour in awake, behaving animals \cite{varley_information-processing_2023}, and has been used to explore the structure and dynamics of organotypic cultures (\cite{ito_extending_2011,ito_large-scale_2014,kajiwara_inhibitory_2021,timme_criticality_2016,faber_computation_2018,newman_revealing_2022}. Here, following \cite{varley_information-processing_2023} we applied the information dynamics framework to spontaneous spiking activity collected from organotypic cultures before, during, and after exposure to the serotonergic psychedelic N,N-dipropyltryptamine (DPT), with the aim of to creating a comprehensive portrait of the way that the psychedelic drug alters information dynamics at the circuit level. 

DPT is a serotonergic psychedelic of the tryptamine class and a close analogue of the more well-known psychedelic, N,N-dimethyltryptamine (DMT, one of the active ingredients in Ayahuasca). DPT has been known to science since the early days of psychedelic research: as early as 1962 it was being explored as a tryptamine analogue of psilocybin \cite{szara_metabolism_1962,faillace_clinical_1967}. By the 1970s, it had become an object of clinical research, being tested as a treatment for alcoholism \cite{grof_dpt_1973}, and later to test if its mystical experience-producing properties might be of use for terminal cancer patients facing the end of their lives \cite{richards_peak_1977}. In the years following the passing of the Controlled Substances Act, scientific and clinical interest in DPT waned, however, it was never criminalized in the United States and it remains unscheduled at the Federal level. Despite its legality, DPT remains much less well-known among the general public than it's more famous siblings such as psilocybin, DMT, mescaline, and LSD. A notable exception to this is it's use by a religious organization based in New York City, The Temple of the True Inner Light, which uses DPT as a religious sacrament \cite{noauthor_temple_nodate}. Despite it's somewhat unusual history and status, pharmacological research has shown it to be a standard serotonergic psychedelic of the tryptamine class, with activity mediated by both the 5-HT$_{\textnormal{2A}}$ and 5-HT$_{\textnormal{1A}}$ receptors, which is typical of the class of drugs in question \cite{li_behavioral_2007,fantegrossi_hallucinogen-like_2008}. Its legal status, and close relationship to more well-known, scheduled drugs, made it an excellent compound for this study.

\section{Results}

\begin{figure}
    \centering
    \includegraphics[width=\textwidth]{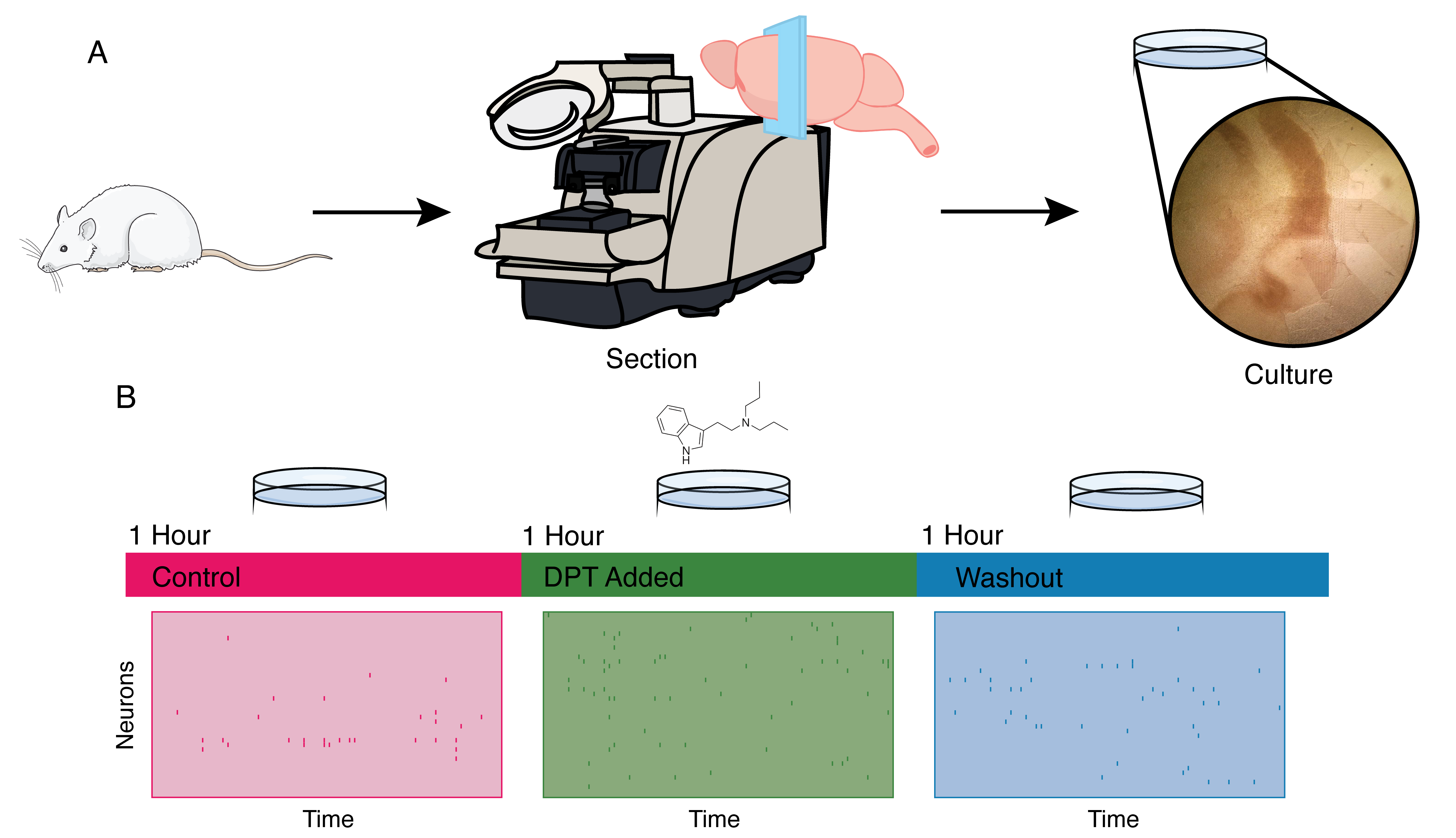}
    \caption{\textbf{Visual explanation of methods.} \textbf{A.} The slices are prepared from cortical tissue of Sprague-Dawley rats, sectioned, and cultured \textit{in vitro} for a period of two weeks. \textbf{B.} Following incubation, cultures were recorded for three hours: one hour before drug administration in a control medium (pink), one hour while being exposed to a 10$\mu$M solution of N,N-DPT (pink), and finally for one hour under control conditions after washout (blue). Example raster plots showing spikes for each condition are showed.}
    \label{fig:explainer}
\end{figure}

\subsection{Summary of Methods}
Here, we will briefly outline the methods and analyses presented in this paper. For more details, see the Materials \& Methods section. To investigate how DPT affects neuronal activity at the meso-scale, we chose to use organotypic cultures of rat somatosensory cortex. Organotypic cultures preserve some of the layered structure that is typical of cortex, yet are compact and easily accessible for fluid changes, as needed in this study. Moreover, these cultures have been shown to display many of the emergent properties reported from recordings of in vivo systems, including wave-like structures \cite{sanchez-vives_cellular_2000}, synchrony \cite{cappaert_spatio-temporal_2009}, gamma oscillations \cite{fisahn_cholinergic_1998}, repeating activity patterns \cite{rolston_precisely_2007}, and neuronal avalanches \cite{beggs_neuronal_2003}. They also display a rich club structure of effective connectivity \cite{ito_large-scale_2014}, as reported in many other neural systems \cite{heuvel_high-cost_2012,dann_uniting_2016,varley_information-processing_2023}. Following prior work \cite{tang_maximum_2008,ito_extending_2011}, organotypic cultures of cortical tissue were taken from 5-day postnatal rats, and after a two week incubation period, spontaneous spiking activity was recorded on a 512-electrode array. Recordings lasted for a three-hour period; in the first hour, the cultures were recorded in their standard environment of cell media. In the second hour, the cultures were exposed to a 10 $\mu$M solution of N,N-dipropyltryptamine at a perfusion rate of 3mL/minute, and in the third hour, the drug was washed out and a subsequent hour of control condition was recorded. We then analyzed how the statistics of population firing activity varied between control, drug, and washout recordings. 

The data was spike-sorted using the \texttt{kilosort3} package \cite{pachitariu_solving_2023}, and analyzed using the \textit{information dynamics} framework \cite{lizier_local_2013} with the aid of the \texttt{IDTxl} package. Information dynamics uses the mathematics of information theory to describe the statistical structure of temporally extended processes, with the ultimate goal of creating an effective model of the distributed ``computations" the system is performing. Due to variability between cultures (such as which specific regions of the somatosensory cortex the initial culture was taken from, precise placement of the electrode array, etc), we aggregated all neurons into a single sample for analysis and do not explore culture-level differences.

The various information-dynamic measures can be grouped into three general categories: first-order measures that describe the dynamics of individual neurons. We considered the Shannon entropy of the spike train (a measure of activity intensity), the active information storage \cite{lizier_local_2012,wibral_local_2014} (a measure of temporal autocorrelation), and the entropy production \cite{roldan_entropy_2012,lynn_broken_2021} (a measure of how time-reversible the dynamics of the elements are). The second set of measures were second order; describing the interactions between pairs of elements. We considered the multivariate transfer entropy \cite{lizier_multivariate_2012,bossomaier_introduction_2016,novelli_large-scale_2019}, a measure of information flow from a ``source" neuron to a ``target" neuron, and for each culture, inferred a multivariate transfer entropy network, after \cite{varley_information-processing_2023}. In addition to the amount of information flow between neurons in bits, we also characterized the local topology of the directed networks with the local clustering coefficient \cite{watts_collective_1998}. The final set of ``higher-order" measures was the statistical synergy between pairs of sources onto a single target (for review, see \cite{newman_revealing_2022}). This serves as a measure of information modification \cite{lizier_towards_2013}, or non-trivial ``computation" in circuits of multiple interacting neurons \cite{timme_high-degree_2016}. Since almost all of the measures returned values spanning multiple orders of magnitude (a typical feature of neural data \cite{buzsaki_log-dynamic_2014}), we log-transformed the values for statistical analysis. Furthermore, since not every neuron was active in every condition, we filtered the neurons and only included those cells that were active in all three conditions; this ensures the validity of the repeated measures design. Finally, information-theoretic measures (AIS, mTE, synergy) were normalized as described in \cite{newman_revealing_2022}, by dividing the measure by the target entropy, which accounts for the variable firing rates that could confound the data. 

Collectively, this suite of measures presents a multi-dimensional perspective on how the serotonergic psychedelic N,N-DPT alters computational dynamics in cortical circuits. We have provided a glossary of reference terms at the end of the manuscript (see Sec. \ref{sec:glossary}), and all the measures are detailed more formally in the Materials and Methods. 

\subsection{First-Order Measures}

Friedman's $\chi^{2}$ found a significant difference in the log-transformed Shannon entropy ($Q\approx174.89$, $p\approx1.06\times10^{-38}$). Post-hoc analysis found that the DPT condition had significantly higher log-transformed entropy ($-2.29 \pm 0.72$) than the both the control condition ($-2.57 \pm 0.99$, $t\approx-12.92$, $p\approx2.06\times10^{-36}$, Cohen's D=-0.32), and the washout condition ($-2.58 \pm 0.9$, $t\approx16.88$, $p\approx6.07\times10^{-59}$, Cohen's D=0.35), but there was no significant difference between the control and washout conditions. This is consistent with whole-brain level findings that serotonergic psychedelics increase the overall entropy of brain activity \cite{carhart-harris_entropic_2014,sarasso_consciousness_2021}. When considering the log-transformed entropy-production (a measure of irreversibility) of the spiketrains, Friedman's test found a significant difference between conditions ($Q\approx80.42$, $p\approx3.44\times10^{-18}$), and posthoc analysis once again found a small, but significantly higher entropy-production (greater irreversibility) production in the DPT condition ($-4.13 \pm 1.39$) when compared to the control condition ($-4.47 \pm 1.57$, $t\approx-8.29$, $p\approx3.71\times10^{-16}$, Cohen's D=-0.23) and the washout condition ($-4.46 \pm 1.45$, $t\approx10.22$, $p\approx2.29\times10^{-23}$, Cohen's D=0.23), but not between control and washout. Recent work on human neuroimaging has found that loss of consciousness is associated with increased reversibly of brain activity \cite{de_la_fuente_temporal_2022,g-guzman_lack_2023}, and so the finding that a psychedelic like DPT is associated with an increase in entropy production suggests that time-reversibility may be a more general marker of conscious states. 

Interestingly, we found no significant differences in the log-transformed active information storage (AIS) between any of the conditions, however, we did find strong, significant differences in the maximum search depth for the embedding lag ($Q\approx348.17$, $p\approx2.49\times10^{-76}$). The maximum search depth can be understood as the ``time-horizon" of the neuron's memory: the maximum distance into the past that still contains information about the immediate future. Post-hoc analysis found that all three conditions were distinct. The DPT condition had the shortest memory (4.04 ms $\pm$ 0.96), lower than both the control ($t\approx8.41$, $p\approx1.59\times10^{-16}$, Cohen's D=0.41) and washout ($t\approx-19.88$, $p\approx4.12\times10^{-73}$, Cohen's D=-1.07) conditions. The control condition was in the middle (4.47 ms $\pm$ 1.14), and significantly lower than the washout condition ($t\approx-8.24$, $p\approx5.9\times10^{-16}$, Cohen's D=-0.41), which had the longest average memory (4.82 ms $\pm$ 0.39). 

Collectively, these results indicate that the dynamics induced by DPT are distinct from the drug-free state: the single-neuron activity in the DPT condition is characterized by higher entropy, less reversible dynamics, as well as a shorter ``memory" in each neuron (although the total AIS was surprisingly unchanged). These results are broadly consistent with what we might expected based on the entropic brain hypothesis.  

\subsection{Network Measures}

After constructing the multivariate transfer entropy network (for details, see Materials \& Methods), we analyzed the structure of directed, pairwise dependencies between neurons. Friedman's test found small, but significant differences between conditions in the log-transformed total information flowing into each neuron ($Q\approx87.35$, $p\approx1.08\times10^{-19}$). Post-hoc analysis found that, once again, there was no significant difference between the control ($-14.79 \pm 5.24$) and washout ($-14.91 \pm 4.21$) conditions, but that the DPT condition had significant less mTE ($-16.15 \pm 3.85$) than either control ($t\approx8.53$, $p\approx6.04\times10^{-17}$, Cohen's D=0.3) or washout ($t\approx-11.46$, $p\approx1.91\times10^{-28}$, Cohen's D=-0.31). 

\begin{figure}
    \centering
    \includegraphics[width=1.0\textwidth]{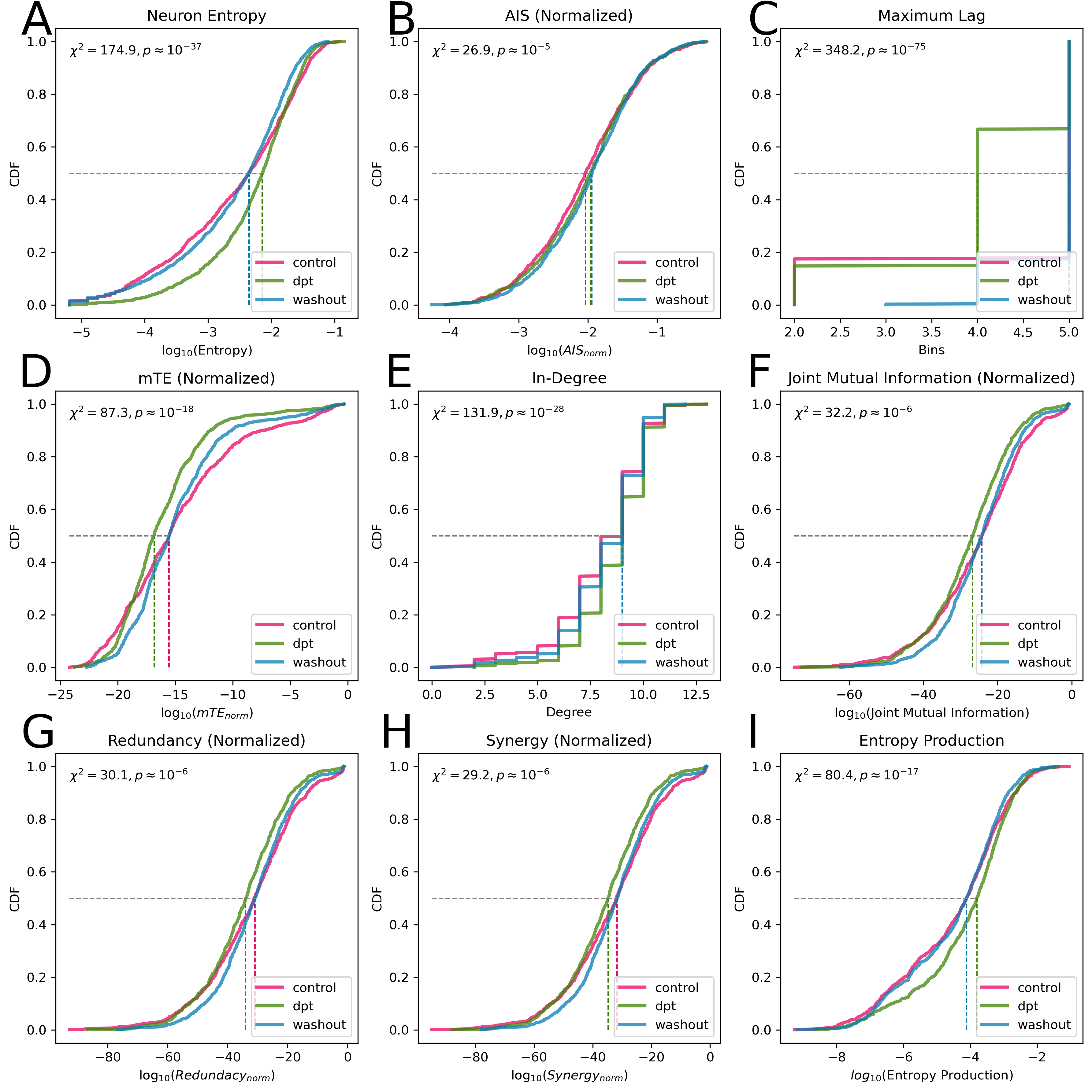}
    \includegraphics[width=1.0\textwidth]{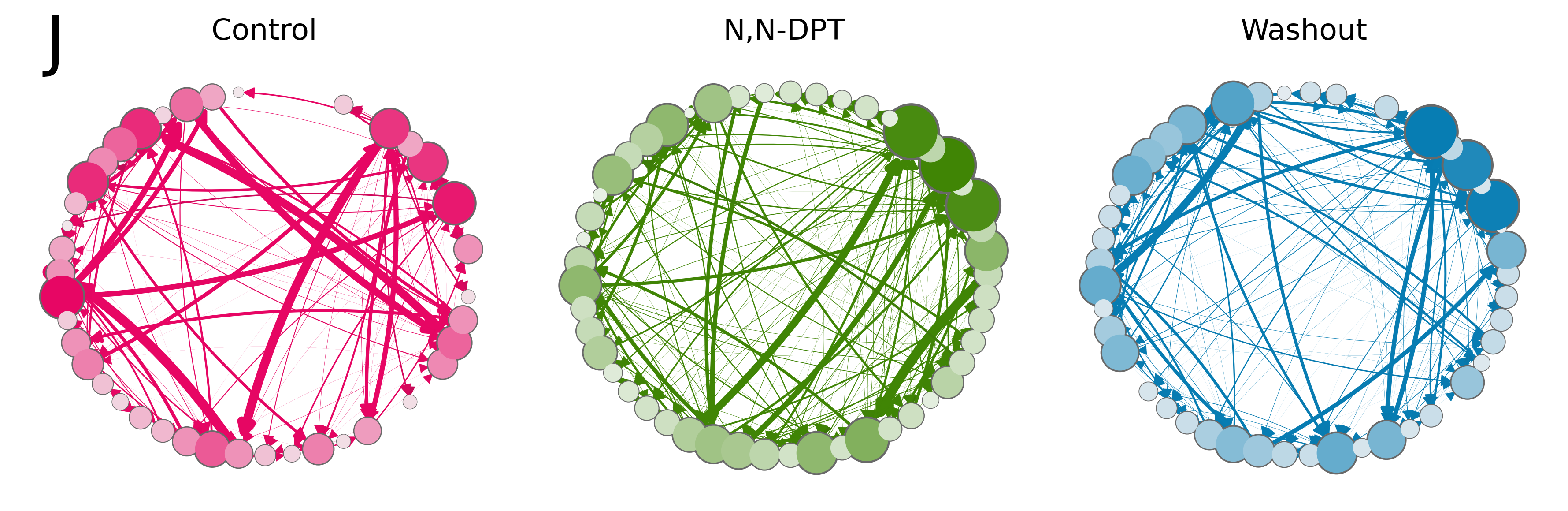}
    \caption{\textbf{A.} A cumulative distribution function (CDF) plots of the individual neuron entropies for the three conditions (pink: control, green: DPT, blue: washout). The Friedman's $\chi^2$ statistic is computed from all three distributions. \textbf{B-I.} CDF plots for the following measures: AIS, AIS maximum lag, total in-coming mTE, in-degree, joint mutual information from two parents onto a single target, redundant information synergistic information, and entropy production. \textbf{J.} Visualization of a representative mTE network from a single culture during all three conditions. Visual inspection shows that the DPT condition has an increased number of weak (thin) edges when compared to the control condition, consistent with the finding that the in-degree of each neuron has increased even as overall information flow decreases.}
    \label{fig:cdfs_graphs}
\end{figure}

Curiously if we consider the discrete in-degree, rather than considering the total information in-flow, we find the opposite pattern: ($Q\approx131.9$, $p\approx2.28\times10^{-29}$). There is no significant difference in in-degree between the control ($8.08 \pm 2.07$ edges) and washout ($8.27 \pm 1.81$ edges) conditions, however, the DPT condition has a significantly greater in-degree ($8.71 \pm 1.6$ edges) than both the control condition ($t\approx-10.37$, $p\approx8.81\times10^{-24}$, Cohen's D=-0.34) and the washout condition ($t\approx10.57$, $p\approx1.33\times10^{-24}$, Cohen's D=0.25). This is curious, as it suggests that, in the DPT condition, there is an increase in low-level connectivity, but that the strength of individual edges is also reduced: a proliferation of weak connections. For visualization of an example network, see Figure \ref{fig:cdfs_graphs}.

This hypothesis is supported by an analysis of local circuit density in the network. Commonly called the ``clustering coefficient" \cite{holland_transitivity_1971,watts_collective_1998}. Briefly, the clustering coefficient gives a measure of local integration: for each neuron, it quantifies how many of that neuron's neighbors are also neighbors (i.e. form closed triangles). Friedman's test found significant differences in the log-transformed clustering coefficient between all conditions ($Q\approx211.73$, $p\approx1.06\times10^{-46}$), and post-hoc analysis found significant differences between all pairs of conditions. The control condition had the lowest log-transformed clustering coefficient (-1.91 $\pm$ 0.61) compared to DPT (-1.7 $\pm$ 0.4, $t\approx-15.22$, $p\approx2.54\times10^{-46}$, Cohen's D=-0.39) and washout (-1.63 $\pm$ 0.39, $t\approx-8.87$, $p\approx4.51\times10^{-18}$, Cohen's D=-0.18), and the washout condition was significantly higher than the DPT condition ($t\approx-8.87$, $p\approx4.51\times10^{-18}$, Cohen's D=-0.18), although note the weak effect size. These results suggest that addition of DPT is associated with an increase in weak, local integration: while the total amount of information coming into each neuron is decreased, more locally clustered weak connections are allowed to open. Curiously, unlike many of the other metrics, this effect persists even after the drug is washed out. 

Collectively, these results challenge simplistic stories such as "increased connectivity" or "decreased connectivity", but rather suggest a more nuanced change in the communicative structure of the network, typified by both an overall decrease in the total information flow, but an increase in the number of weak open connections. 

\subsection{Higher-Order Statistical Synergy}

When considering higher-order information integration (statistical synergy), we found weak, but significant patterns consistent with prior results. Friedman's test on the log-transformed normalized synergy found significant differences between the conditions ($Q\approx29.2$, $p\approx4.57\times10^{-7}$). Post-hoc analysis found no significant difference between the control ($-33.21 \pm 15.84$) and washout conditions ($-32.24 \pm 13.01$), but a weak, significant decrease in log-transformed synergy in the DPT condition ($-35.54 \pm 13.64$) compared to both control ($t\approx4.35$, $p\approx1.51\times10^{-5}$, Cohen's D=0.16) and washout conditions ($t\approx-7.43$, $p\approx2.73\times10^{-13}$, Cohen's D=-0.25). These results tentatively suggest that, when exposed to DPT, the individual neurons are ``integrating" less information from pairs of inputs then they ordinarily would. This finding was unexpected, as previous research has found that loss of synergy is generally associated with decreased conscious awareness \cite{luppi_reduced_2023,luppi_what_2021}, although this prior work has been done exclusively at the whole-brain level. We stress that these are tentative results for two reasons, however: the first is that different redundancy functions or formulations of the PID may return different synergies \cite{kolchinsky_novel_2022}, and the second is that we only considered the case of two parents and a single target: higher-order combinations may show quantitatively different patterns of information integration, although such an analysis is beyond the scope of this project. 

\section{Discussion}

In this paper, we have described how the serotonergic psychedelic N,N-dipropyltryptamine (DPT) alters the statistics of information dynamics in organotypic cultures before, during, and after drug exposure. We found that concentrations of 10 $\mu$M DPT induced a transient dynamic characterized by increased entropy of single neuron activity, reduced strong connections between neurons, but simultaneously, a proliferation of weak connections. We found that higher-order statistical synergy was decreased, but the temporal irreversibility of neural activity was increased. Collectively, these results paint a complex picture of the effects of DPT on neural circuit dynamics. The decrease in strong connections and reduction in synergistic processing could be described as ``disintegration" of the system: in both cases, smaller proportion of the uncertainty about the future activity of the target neurons can be resolved by learning about other parts of the system. Conversely, however, the increase in in-degree (indicating a growth in weak connections) suggests that this is not the entire story: more channels of information flow may be opening, they are just weaker in nature.  

These results are broadly consistent with prior results from whole-brain neuroimaging. The increase in regional entropy is well-documented enough to form the core of the entropic brain hypothesis \cite{carhart-harris_entropic_2014,carhart-harris_entropic_2018} (although for a dissenting opinion, see \cite{mcculloch_navigating_2023}). Similarly, bivariate transfer entropy analysis of MEG data from humans under the influence of LSD and psilocybin found decreased effective connectivity \cite{barnett_decreased_2020}. To the best of our knowledge, at the time time of writing, there have been no published analyses of how psychedelic drugs impact temporal reversibility or statistical synergy (although Mediano reports that a closely related measure, integrated information, $\Phi$, surprisingly decreases under LSD or psilocybin in a manner somewhat similar to sleep \cite{mediano_integrated_2019}).

The finding that DPT induces an increase in weak connections may provide insights into the documented ability of tryptamine psychedelics to induce neuroplasticity in neuronal networks. \textit{In vitro} work has found that exposure to drugs such as LSD and psilocybin produces increased dendritic arborization and synaptogenesis \cite{ly_psychedelics_2018,de_la_fuente_revenga_prolonged_2021}. A naive Hebbian model might suggest that it is the increased information flow between previously dis-connected neurons that might drive the emergence of new connections, although we should stress that the transfer entropy network inference algorithm does not claim to recover purely synaptic connections. Future work that can combine spontaneous activity recording with biological analysis of neuroplasticity may be able to explore the connection more directly. 

Curiously, despite the consistency with macro-scale imaging analyses, the finding that DPT increased the entropy of spontaneous firing activity relative to the control and washout conditions conflicts with two prior cellular-level studies, both of which found that the psychedelic 2,5-Dimethoxy-4-iodoamphetamine (DOI) had an inhibitory effect on spiking activity \cite{wood_disruption_2012,rangel-barajas_dysregulated_2017}. One possible explanation for this discrepancy is the different pharmacological profiles of the two drugs: DOI is a substituted amphetamine, while DPT is of the tryptamine class, and they have distinct binding profiles. Another possibility is the difference between \textit{in-vivo} and \textit{in-vitro} studies. Given the overall paucity of research on the circuit-level effects of psychedelic drugs on neural dynamics, further studies will hopefully shed considerable light on these questions.

This study has some limitations that are worth discussing. The most significant is the small absolute number of recordings (11), which makes culture-to-culture comparisons weak (in contrast to neuron- and circuit-level analyses, which are highly powered). The cultures themselves have no behavior or consciousness to speak of, and so the insights that can be gleaned from them about the phenomenological nature of the psychedelic state are limited. The cultures themselves are taken from the dorsal cortex near the somatomotor areas, however the precise placement of the electrodes varies, which means there is unavoidable heterogeneity with respect to which neurons are being sampled and what layers are represented. Future replications with larger N, and possibly in behaving animal models will go a long way to addressing these concerns. Recent developments in multi-layer imaging from animal cortex \cite{kajiwara_inhibitory_2021}, or machine-learning based cell-type classification \cite{lee_non-linear_nodate}, may augment future studies in this vein. Finally, this study compares DPT to an empty vehicle solution (DMSO), and since DPT is a relatively promiscuous ligand (binding to many different serotonin receptors), it is impossible to attribute the observed effects to any single receptor. 

These results should be seen as a first step towards understanding the effects of psychedelics on circuit-level information-processing dynamics. The limitations discussed above suggest natural subsequent studies, including using invasive recordings from behaving animals (where placement of the array can be controlled), studying the dose-response curves with respect to measures like neural entropy, and finally, increasing the population size to improve statistical power. However, despite the limitations, we suggest that this study has provided key insights into the computational effects of psychedelics on meso-scale brain activity.

\section{Conclusions}

In this study, we showed that the serotonergic psychedelic N,N-DPT disrupts information-processing dynamics of cortical tissue in \textit{in vitro} organotypic cultures, with some disruptions appearing to be reversible, while others persist post-exposure. The psychedelic increased the entropy of spontaneous neural firing activity, while decreasing the temporal reversibility, and altered the connectivity patterns of neural circuits: reducing the overall information flow coming into each neuron, but increasing the total number of significant connections. These different effects present a nuanced picture, largely irreducible to simple stories of "increasing integration" or "decreasing integration", and instead point to a rich area of future work more carefully characterizing the effects of psychedelics on information-processing, and computational, dynamics in the brain. 

\section*{Glossary}
\label{sec:glossary}
Here we will provide a brief reference of the various information-theoretic and graph-theoretic measures described here. For readers interested in finer detail, see the Materials and Methods section and the references therein. 

\begin{itemize}
    \item[] \textbf{Entropy:} A measure of uncertainty about the outcome of a random draw from a distribution. Formally: $H(X) = -\sum_{x\in\mathcal{X}}P(x)\log P(x)$, where $\mathcal{X}$ is the support set of $X$.
    \item[] \textbf{Support Set:} For a random variable $X$, the support set of $X$, denoted as $\mathcal{X}$ is the set of all possible states $X$ can adopt. 
    \item[] \textbf{Mutual Information:} The amount of uncertainty about a variable $X$ reduced upon learning the state of some other variable, $Y$. Formally: $I(X;Y) = H(X) - H(X|Y)$, where $H(X|Y)$ is the conditional entropy of $X$ given $Y$. 
    \item[] \textbf{Active Information Storage:} For a temporally extend process $X$, the AIS is the information about the state of $X_t$ at time $t$ disclosed by the past. Formally $AIS(X_t)=I(X_{past};X_t)$.
    \item[] \textbf{Maximum lag:} The maximum distance in the past accounted for when computing the AIS. 
    \item[] \textbf{Transfer Entropy:} The information that the past of one variable $X$ discloses about the next state of a target variable $Y$, conditioned on $Y$'s own past. Formally: $TE(X\to Y)=I(X_{past};Y_t|Y_{past}).$
    \item[] \textbf{Multivariate Transfer Entropy:} The information that the past of one variable \textbf{X} discloses about the next state of a target variable $Y$ conditioned on the past of all other parents of $Y$. Formally: $I(X_{past};Y_{t}|Y_{paste}, \textbf{Z}_{past}$, where \textbf{Z} is the set of all parents of $Y$ excluding $X$.
    \item[] \textbf{Partial Information Decomposition:} A technique for decomposing the information that two sources ($X_1$ and $X_2$) disclose about a single target $Y$ into redundant, unique, and synergistic components. 
    \item[] \textbf{Redundant Information:} The information about a target $Y$ that could be learned by observing either $X_1$ alone or $X_2$ alone. 
    \item[] \textbf{Unique Information:} The information about a target $Y$ that can only be learned by observing $X_i$.
    \item[] \textbf{Synergistic Information:} The information about a target $Y$ that can only be learned by observing the joint state of $X_1$ and $X_2$ simultaneously.
    \item[] \textbf{Entropy Production:} A measure of the time-reversibility of a temporal process. Formally: $D_{KL}(\overrightarrow{X}||\overleftarrow{X})$.
    \item[] \textbf{Kullback-Leibler Divergence:} The information gained when updating one's belief from a prior distribution to a posterior distribution. Formally: $D_{KL}(P||Q) = \sum_{x\in\mathcal{X}}P(x)\log P(x)/Q(x)$ where $P$ and $Q$ are both probability distributions on the support set $\mathcal{X}$.
    \item[] \textbf{In-Degree:} The number of in-coming edges to a node in a network. 
    \item[] \textbf{Out-Degree:} The number of out-going edges from a node in a network. 
    \item[] \textbf{Local Clustering Coefficient:} A measure of how many triangles a given node participates in relative to the total number of triangles it could possibly participate in given it's degree. 
    \end{itemize}

\section{Materials \& Methods}

\subsection{Organotypic Culture Preparation, Data Collection \& Preprocessing}

Organotypic cultures were prepared according to the methods described in \cite{tang_maximum_2008,ito_extending_2011}. Briefly, we used Sprague-Dawley strain postnatal rats which were an average age of five days old. These animals were approved by the Indiana University Animal Care and Use Committee, and all proper protocols for animal care were followed. The overall procedure involved extracting their brains and slicing them in the coronal plane using a vibrotome to achieve a thickness of 400 $\mu$m. After this process, the slices were placed in trays with culture medium in an incubator for a time period between two to four weeks. The culture medium in the trays was replaced by half every three days. The composition of the culture medium is as following: 1L Minimum Essential Medium (Sigma-Aldrich), 500mL Hank’s balanced salt solution (Sigma-Aldrich), 500mL of heat inactivated horse serum (Sigma-Aldrich), 2mL PSN antibiotic mixture, and 10mL L-Glutamine.

All animal tissue samples were prepared according to guidelines from the National Institutes of Health and all animal use procedures were approved by the Indiana University Animal Care and Use Committee (IUCAC). After 2-4 weeks of maturation, cultures were recorded on a 512-microelectrode array, with 5 micron diameter electrodes arranged in a triangular lattice with an inter-electrode distance of 60 $\mu$m \cite{litke_what_2004}. Data were sampled at a high temporal resolution of 50 $\mu$s.

Each culture was recorded from for three hours. The first hour was the control condition; spontaneous activity was recorded under normal conditions. A ``placebo" of [X] $\mu$L of empty DMSO vehicle were added to the culture media. Following the control hour, the irrigation system was flushed, and a second batch of culture medium, containing 10 $\mu$M N,N-DPT solution in DMSO (Cayman Chemical Company) was introduced. Cultures were recorded from for another hour (the drug condition), before the system was again flushed and the original, drug-free media was re-introduced. Recordings were stopped during media turn-over to avoid artifacts. 

Following recording, the three one-hour datasets were appended, and spike-sorting was done using the \texttt{kilosort3} software package \cite{pachitariu_solving_2023}, in a Python3.7 environment. Following spike-sorting, the resulting rasters were re-binned to 1 ms frames. Rasters were excluded from analysis if they contained less than 30 neurons, resulting in a final count of 11 viable datasets. 

\subsection{Information Dynamics \& Network Inference}

Information dynamics is a quantitative framework used to analyze how the elements of a complex system interact and collectively ``compute" the future trajectory of a system \cite{lizier_local_2013}. By drawing on analogy with digital computation, the information dynamics framework breaks ``computation" in complex systems down into a set of distinct dynamical features, including information storage (analogous to memory, or autocorrelation), information flow or transfer, and information modification or ``integration".

For an element $X$ in a stochastic dynamical system, the simplest measure of information structure is the Shannon entropy of that element: how uncertain are we, as observers, about the state $X$ will adopt at time $t$? Formally:

\begin{equation}
    H(X_t) = -\sum_{x\in\mathcal{X}}P(x)\log_2P(x)
\end{equation}

Where $\mathcal{X}$ is the support set of $X$ and $P(x)$ is the probability of observing that $X=x$. The Shannon entropy has no notion of dynamics, however: it assumes that every time $t$, $X$ is randomly selected it's state from $\mathcal{X}$ according to $P(x)$. 

\subsubsection{Active Information Storage}

The simplest measure of information dynamics is the active information storage, which quantifies how much the past state of $X$ constrains the possible next state $X_t$:

\begin{eqnarray}
    \label{eq:ais}
    AIS(X) &=& I(X_{past};X_t) \\
    &=& H(X_t) - H(X_t|X_{past}) \nonumber
\end{eqnarray}

Where $X_{past}$ refers to a potentially multidimensional embedding of the past states of $X$. 

We can re-write Eq. \ref{eq:ais} as a kind of ``information regression" that details how information about $X$'s next state is distributed over time \cite{lizier_multivariate_2012}:

\begin{equation}
    H(X_t) = AIS(X) + H_{\mu}(X)
\end{equation}

Here $H_{\mu}(X)$ is the conditional entropy rate $H(X_t|X_{past})$: all that uncertainty about $X_t$ that is not resolved by learning the past of $X$. 

For each neuron in each recording, for each condition, we inferred the AIS using a non-uniform embedding algorithm provided by the \texttt{IDTxl} package \cite{wollstadt_idtxl_2019}. Briefly, the non-uniform embedding procedure iterates through lags $1..\tau_{\max}$ (inclusive) and tests whether the addition of each subsequent lag significantly increases the AIS, conditional on all previously selected lags, up to some maximal lag $\tau_{\max}$. For more details, see \cite{faes_information-based_2011} and the \texttt{IDTxl} documentation. Here $\tau_{\max}$ was chosen to be 5 bins, and 1000 shuffled nulls were used for null-hypothesis significance testing. To control for the effects of variable firing rates, we report the normalized active information storage: $AIS(X)/H(X_t)$.

\subsubsection{Multivariate Transfer Entropy}

The AIS quantifies how much information the past of a single element discloses about it's own future (the amount of information ``stored" in $X$). To quantify how much information ``flows" from one element to another, we must measure how the past of other elements of the system constrains $X_t$. This is done with the multivariate transfer entropy \cite{schreiber_measuring_2000,novelli_large-scale_2019}. For a set of parent elements $\textbf{Z}$, we can quantify how much information the past of \textbf{Z} discloses about the next state of $X$ with the conditional mutual information:

\begin{equation}
    mTE(\textbf{Z}\to X) = I(\textbf{Z}_{past} ; X_t|X_{past})
\end{equation}

In the context of the infomration regression, we now have:

\begin{equation}
    H(X_t) = AIS(X) + mTE(\textbf{Z}\to X) + H_{\mu}(X)
\end{equation}

Where $H_{\mu}$ is now given by $H(X_t|X_{past},\textbf{Z}_{past})$. The $mTE$ is appealing in that it accounts for potentially higher-order synergies between multiple $Z_i,X_j\in\textbf{Z}$, as well as not double-counting redundancies as the bivariate transfer entropy does \cite{bossomaier_introduction_2016}. The full $mTE(\textbf{Z}\to X)$ is a multivariate measure, more naturally applicable to hypergraphs than bivariate networks, however, a bivariate network that still accounts for redundancies and synergies can be recovered by defining the weight of each directed edge as $I(Y_{past}:X_t|\textbf{Z}^{-Y}_{past}, X_{past})$, where $\textbf{Z}^{-Y}$ refers to the set of all $Z_i\in\textbf{Z}$ excluding $Y$.

For large systems with finite datasets, it is impossible to account for all possible parents, as well as all possible lags. Here, we used the \texttt{IDTxl} package \cite{wollstadt_idtxl_2019} to implement a modified version of the algorithm described in \cite{novelli_large-scale_2019}. \texttt{IDTxl} implements a greedy search, coupled with extensive null hypothesis surrogate testing to infer an optimal parent set \textbf{Z} and the embedding for both $X_{past}$ and $\textbf{Z}_{past}$, however, the runtimes can still be excessive: the time complexity for a full network inference is $O(N^{2}\times d\times \tau_{max}\times S)$, where $N$ is the number of neurons in the network, d is the eventual average in-degree of each neuron, $\tau_{\max}$ is the maximum search depth, and $S$ is the number of surrogates \cite{wibral_m_runtimes_2021}. Given limitations in available computing resources, we first pre-filtered the set of prospective parents for each target by removing any neurons that did not have any significant bivariate transfer entropy onto the target over a range of 1..30 bins of lags for the source and five ms bins of lag for the target. Significance was tested using the analytic null estimator \cite{barnett_transfer_2012} as implemented by the \texttt{JIDT} \cite{lizier_jidt_2014}. The results of the analysis were fed into the \texttt{IDTxl} mTE estimator. Following prior work on transfer entropy network inference in neural cultures \cite{ito_extending_2011,shimono_functional_2015,nigam_rich-club_2016,timme_multiplex_2014}, we constrained the multivariate transfer entropy inference to only consider one bin of source history, fixed by the lag that maximized the significant bivariate transfer entropy. The parent set \textbf{Z} for each neuron, in each culture, in each condition was inferred in parallel (requiring approximately 5,000 unique optimizations), and significance testing was done using null distributions of 250 circularly-shifted surrogates. The circular shift was chosen to preserve the autocorrelation of each neuron. To control for the effects of variable firing rate, we report the normalized multivariate transfer entropy: $mTE(Y\to X|\textbf{Z}^{-Y})/H(X_t)$.

\subsubsection{Partial Information Decomposition \& Synergy}

The final information dynamic we explored is information modification \cite{lizier_local_2013}, sometimes also referred to as information integration. Information modification has been associated with ``computation" in neural systems previously \cite{newman_revealing_2022}, and refers to novel information generated when a single neuron's future is constrained by the joint state of multiple inputs simultaneously \cite{lizier_towards_2013}. Following previous work \cite{timme_high-degree_2016,newman_revealing_2022,varley_information-processing_2023}, we operationalized information modification with the statistical synergy, as computed using the partial information decomposition (PID) framework \cite{williams_nonnegative_2010}.

Since it's development by Williams and Beer in 2012, the PID framework has been widely applied across a variety of fields, including neuroscience \cite{newman_revealing_2022,luppi_synergistic_2022}, clinical care reserach \cite{luppi_reduced_2023}, sociology \cite{varley_untangling_2022}, climatology \cite{goodwell_temporal_2017}, machine learning \cite{ehrlich_measure_2023}, as well as to philosophical questions such as ``emergence" \cite{mediano_greater_2022,varley_emergence_2022} and consciousness \cite{luppi_what_2021}. Briefly, the PID provides a scaffold by which the information that multiple sources disclose about a target can be decomposed into non-overlapping ``atomic" components of information. Consider the case where two parent neurons $Y_1, Y_2$ disclose information about a target neuron $X$. The total information that both parents disclose about the target can be quantified with the joint mutual information: $I(Y_1,Y_2;X)$, however this is a lump sum measure and treats $Y_1$ and $Y_2$ as a coarse-grained macro-variable and reveals nothing about how the information about $X$ is distributed over the $Y_i$'s. The PID solves this issue by decomposing:

\begin{equation}
    I(Y_1,Y_2;X) = Red(Y_1,Y_2;X) + Unq(Y_1;X/Y_2) + Unq(Y_2;X/Y_1) + Syn(Y_1,Y_2;X)
\end{equation}

The term $Red(Y_1,Y_2;X)$ is the redundant information about $X$ that could be learned by learning either the state of $Y_1$ alone or the state of $Y_2$ alone. The term $Unq(Y_i;X/Y_j)$ is the unique information about $X$ that can only be learned by observing $Y_i$. The final term, $Syn(Y_1,Y_2;X)$ is the synergistic information about $X$ that can \textit{only} be learned when both the states of $Y_1$ and $Y_2$ are observed simultaneously. We can also decompose the marginal mutual informations:

\begin{align}
    I(Y_1;X) = Red(Y_1,Y_2;X) + Unq(Y_1;X/Y_2) \\
    I(Y_2;X) = Red(Y_1,Y_2;X) + Unq(Y_2;X/Y_1)
\end{align}

The result is an under-determined system of three equations and four unknown values (the redundant, synergistic, and two unique information atoms): if any one term is computed, the remaining three can be solved ``for free." Here we used the $I_{BROJA}$ measure of unique information \cite{bertschinger_quantifying_2014}, as it guarantees a non-negative decomposition. For each network, in each condition, we computed the bivariate PID for every instance of the two-parent/single target motif with the \texttt{BROJA-2PID} package \cite{makkeh_broja-2pid_2018}, as provided by the \texttt{IDTxl} package \cite{wollstadt_idtxl_2019}. For each parent, we used the same optimal lag as was used in the mTE network inference. To control for variable firing rates, we report the normalized synergy $Syn(Y_1,Y_2;X)/H(X_t)$.

\subsubsection{Reversibility \& Entropy Production}

To assess whether DPT altered the temporal reversibility of cortical activity, we computed the entropy production in the neuron level time series of spiking activity. Based on \cite{roldan_entropy_2012,lynn_broken_2021}, we estimated the entropy production with the Kullback-Leibler divergence between the forward- and reverse-time spike trains for each neuron: $D_{KL}(\overrightarrow{X}||\overleftarrow{X})$. For a discrete random variable $X$ that transitions from state $x_i$ to state $x_j$ according to a stationary transition probability matrix $P(x_i\to x_j)$, the entropy production is given by:

\begin{equation}
    D_{KL}(\overrightarrow{X}||\overleftarrow{X}) := \sum_{x_i,x_j\in\mathcal{X}}P(x_i\to x_j)\log\bigg(\frac{P(x_i\to x_j)}{P(x_j\to x_i)} \bigg)
\end{equation}

If $P(x_i\to x_j)=P(x_j\to x_i)$ for all $x_i$, $x_j$, then the system is said to obey ``detailed balance" and is at thermodynamic equilibrium: there is no ``flow of time" from the perspective of the system: the flow from past to future and from future to past are indistinguishable. On the contrary, if $P(x_i\to x_j)\not=P(x_j\to x_i)$, then the system has broken detailed balance and is operating far from equilibrium \cite{lynn_broken_2021}. 

To ensure that the state spaces were large enough to capture rich temporal dynamics, we used a lossless coarse-graining procedure on each neuron's spike train: the time series was compressed into non-overlapping, successive 5 ms bins. Each macro-frame could be in one of thirty two possible states and we computed the TPM from the sequence of successive macro-frames. To satisfy the constraints of the $D_{KL}$, we only included transitions where both the transition $x_i\to x_j$ and $x_j\to x_i$ were observed.

\subsubsection{Clustering Coefficient}

The local clustering coefficient \cite{holland_transitivity_1971,watts_collective_1998} for each node in each network was computed using the \texttt{clustering()} function from the \texttt{NetworkX} package \cite{hagberg_exploring_2008}. Briefly, the local clustering coefficient of a node quantifies was proportion of that nodes neighbors are themselves connected. a high value of the coefficient indicates greater local integration. 

\subsection{Software, Code, \& Data Availability Statement}

Data analysis scripts will be provided as supplementary materials for this study. The spike-sorted data will be deposited in the CRCNS repository. Raw, unprocessed data is available from the authors upon request. 

\section{Acknowledgements}
T.F.V. and M.P. are supported by the NSF-NRT grant 1735095, Interdisciplinary Training in Complex Networks and Systems. M.P. is also supported by the NSF GRFP. J.M.B. is supported by Expeditions: Mind in Vitro- Computing with Living Neurons National Science Foundation 2123781 subcontract to J.M.B. This work was supported by the Source Research Foundation.


\end{document}